\documentclass[aps,prl,superscriptaddress,showpacs,preprintnumbers,twocolumn]{revtex4}
\usepackage{epsf,amssymb,amsmath,latexsym}
\usepackage{graphicx}
\newcommand{\be}{\begin{eqnarray}}
\newcommand{\ee}{\end{eqnarray}}
\def\ben{\begin{equation}}
\def\een{\end{equation}}
\def\bena{\begin{eqnarray}}
\def\eena{\end{eqnarray}}

\renewcommand{\a}{\hat a}

\begin{document}

\preprint{DAMTP-2006-35}





%
\title{Two-dimensional Quantum Black Holes, Branes in BTZ and Holography}

\vspace{.3in}

\author{Cristiano Germani}
\email{C.Germani@damtp.cam.ac.uk}\affiliation{D.A.M.T.P., Centre
for Mathematical Sciences, University of Cambridge, Wilberforce
road, Cambridge CB3 0WA, England}

\author{Giovanni Paolo Procopio}
\email{G.P.Procopio@damtp.cam.ac.uk}\affiliation{D.A.M.T.P., Centre
for Mathematical Sciences, University of Cambridge, Wilberforce
road, Cambridge CB3 0WA, England}


\vskip.3in

\begin{abstract}
We solve semiclassical Einstein equations in two dimensions with a 
massive source and we find a static, thermodynamically stable, quantum black hole solution in the
Hartle-Hawking vacuum state. We then study the black hole geometry generated by a
boundary mass sitting on a non-zero tension 1-brane embedded in a three-dimensional
BTZ black hole. We show that the two geometries coincide and we extract, using holographic
relations, information about the CFT living on the 1-brane. Finally, we show that
the quantum black hole has the same temperature of the bulk BTZ, as expected from
the holographic principle.
\end{abstract}

\pacs{04.60.Kz, 04.70.Dy, 11.25.Tq}


\maketitle

\section{Introduction}

In the framework of the proposed duality between gravity on $(d+1)$-dimensional Anti de Sitter ($AdS_{d+1}$) spaces
and d-dimensional conformal field theory ($CFT_d$) living on the $AdS_{d+1}$ boundary, firstly proposed by
Maldacena \cite{Maldacena},
Witten \cite{Witten} suggested a duality between Schwarzschild-AdS black holes ($SAdS$) and a conformal field
theory ($CFT$) at high temperature ($TCFT$) on the $SAdS$ boundary. This idea can be naively understood thinking
that very massive black holes, although stable, emit a black body radiation \cite{Hawbh}. However, as the black
body spectrum does not carry information, the Hawking mechanism is usually associated to a non unitary process
\cite{predictability}. A $TCFT$ is a unitary theory therefore $SAdS$ black holes cannot be fully dual to a
$TCFT$. Indeed this is the case \cite{BR}. We can easily understand why by considering the three-dimensional BTZ
black hole \cite{BTZ}.

The metric for the BTZ black hole is
\begin{equation}\label{btz}
ds^2=-F(r)dt^2+\frac{dr^2}{F(r)}+r^2d\theta^2\ ,
\end{equation}
where
\begin{equation} \label{F}
F(r)=\frac{r^2}{L^2}-m\ , \quad \theta \equiv \theta +2 \pi\ , \quad \ 0\leq r<\infty\ .
\end{equation}
The horizon of this black hole is in $r_h=\sqrt{m}L$, $m$ is its mass and $L$ the AdS length. This spacetime is a
solution of Einstein equations
\begin{equation*}
    R_{\mu \nu}-\frac{1}{2}R g_{\mu \nu}=\frac{1}{L^2}g_{\mu \nu} \ .
\end{equation*}

For a scalar field $\Phi$ propagating in this background the action is given by
\begin{equation*}
{\cal A}(\Phi)=\int_{\mbox{\tiny BTZ}} \sqrt{-g} | d\Phi |^2 +
\int_{\partial\mbox{\tiny BTZ}}\sqrt{-h} \Phi_0 d\Phi\Big|_{\partial\mbox{\tiny{BTZ}}}\ ,
\end{equation*}
where the first integral represent the bulk action and the second the boundary action, $h$ and $\Phi_0$ are the
induced metric and the value of the scalar field on the BTZ boundary (here denoted by $\partial\mbox{BTZ}$)
respectively. The $AdS/CFT$ correspondence relates the above boundary action to the partition function of a
scalar operator in the dual conformal field theory, in this case a $TCFT$, in the following formal way
\be
\nonumber \langle \exp{\int_{\partial\mbox{\tiny BTZ}}\Phi_0{\cal O}}\rangle_{TCFT}={\cal A}_b(\Phi_0)\ .
\ee
$\Phi_0$ now represents the source of a scalar operator ${\cal O}$ of conformal dimension
$\Delta=(1+\sqrt{1+\mu^2L^2})/2$, where $\mu$ are the Kaluza-Klain masses of the solutions of
$\square_{\partial\mbox{\tiny BTZ}} \Phi_0=-\mu^2\Phi_0$ \cite{vakkuri}. With this prescription, one can
calculate correlation functions of the scalar operator ${\cal O}$ in the usual way. For the two-point
correlation function we have
\begin{equation} \label{vaccuri1}
\begin{split}
&\langle {\cal O}(x^a){\cal O}(x_0^a)\rangle_{TCFT}=
\\ &=\frac{\delta^2}{\delta\Phi_0(x^a)\delta\Phi_0(x_0^b)}
\langle\exp{\int_{\partial\mbox{\tiny BTZ}} \Phi_0{\cal
O}}\rangle_{TCFT}\Big|_{\Phi_0=0}= \\
&=\frac{\delta^2}{\delta\Phi_0(x^a)\delta\Phi_0(x_0^b)}{\cal
A}_b(\Phi_0)\Big|_{\Phi_0=0}=G(x^a,x_0^a)\ ,
\end{split}
\end{equation}
where $x^a, x_0^a$ are the boundary coordinates and $G(x^a,x_0^a)$ is called the bulk to boundary correlator. At
very large time \eqref{vaccuri1} is given by \cite{vakkuri}
\begin{eqnarray}\label{corr}
G(x^a,x_0^a)\sim e^{-2\sqrt{m}\Delta(t+t_0)/L}\ .
\end{eqnarray}
We note that the correlator (\ref{corr}) is exponentially decaying
and this is a signal that information has been lost for this
background. Indeed, for a conformal field theory at finite
temperature, we expect that its two point correlation functions
oscillate in a quasi-periodic manner with the quasi-periodicity
dictated by the Poincarr\`{e} recurrence
\cite{BR,Birmingham:2002ph,Solodukhin:2005qy}. To solve this puzzle,
Maldacena in \cite{eternal} and then Hawking in \cite{Haw},
suggested that the correct bulk to boundary operator describing the
boundary theory should be of the form \be G=\sum_{{\cal
M}_i}G_i({\cal M}_i)\ , \ee where the sum is over all the possible
topologies ${\cal M}_i$ satisfying the same boundary conditions.
This is very reminiscent of the path integral sum over histories.
Taking account of only the BTZ background is similar to coarse
graining the phase space in the Feynman path integral. In this way
the apparent information loss as seen from the black hole
perspective is very similar to the information loss in the collapse
of wave functions in ordinal quantum mechanics.

However bulk black hole solutions might be very useful to understand the behavior of semiclassical black holes.
In fact if we consider a conformal field theory at finite temperature with a UV cutoff (equivalent to a coarse
grained process), this will induce a classical gravitational field and will not necessarily need to be unitary.
Obviously this approximation will break down at the quantum gravity regime where we should restore the
unitarity. In order to obtain this theory as a boundary of some asymptotically $AdS$ spacetime we need to
truncate the boundary from spatial infinity to a finite point, as suggested by \cite{em,tan} (see \cite{porran}
for the zero temperature case) to holographically explain the no-go theorem of \cite{cri}. The $d$-dimensional
gravitational effective theory on the boundary will be governed by the semiclassical Einstein equations
\be\label{semic} R_{ab}-\frac{1}{2}Rg_{ab}+\Lambda g_{ab}=k^{2}\langle T_{ab}\rangle\ . \ee $R_{ab}$ is the
Ricci tensor projected onto the brane, $k^{-2}=M_*^{d-2}$ with $M_*$ being the fundamental mass scale (for $d=2$
$k^2=1$), and $\langle T_{ab}\rangle$ is the expectation value of the energy momentum tensor of the conformal
field theory with a UV cutoff $\sigma$ related with the boundary tension, i.e. the trace of the boundary
extrinsic curvature $K^\alpha{}_\alpha=K$. The duality has been checked for the case of three-dimensional brane
black holes embedded in a four-dimensional bulk spacetime \cite{em} and evidences were given in the case of a
four dimensional black hole formation in a five-dimensional embedding \cite{cri2}.

If the bulk is a three-dimensional space, as in the BTZ case, the boundary is a two dimensional surface and
(\ref{semic}) reduces to \cite{Balbinot:1984xf,san}
\be \label{se2}
\Lambda g_{ab}=\langle T_{ab}\rangle\ .
\ee
This case is very interesting as the dynamics
of the gravitational field is of purely quantum origin. Note also that the presence of a cosmological constant
is necessary because of the trace anomaly $\langle T^a{}_{a}\rangle\neq 0$.

In this paper we will find a solution to \eqref{se2} in the presence of a boundary mass. We will find a static,
thermodynamically stable quantum black hole solution in thermal equilibrium in the Hartle-Hawking state.
We will then consider a three-dimensional braneworld model with a BTZ bulk and we will show how our
quantum solution can also be obtained by slicing this three-dimensional spacetime with a non-zero tension, asymptotically
$AdS$,  1-brane. In this way we prove that the conjectured duality between classical bulk black holes and quantum
brane black holes of \cite{em,tan} applies to our case and we extract, using holographic relations, information about the
CFT living on the 1-brane.

\section{Quantum black hole in two dimensions}

We start by considering the two dimensional action
\be \nonumber
I=\frac{1}{2}\int d^2x (R-2\Lambda)\sqrt{-h}+\int d^2x \sqrt{-h}{\cal
L}_{CFT}\ ,
\ee
where ${\cal L}_{CFT}$ is the Lagrangian of a conformal field theory. At this action we add a
Gibbons-Hawking term \cite{GH}
\be\label{boundary}
I_b=-\int dx^a b_a (K+{{\cal L}_b})\ ,
\ee
where $b^a$ is the normal to the boundary, ${\cal L}_b$
is a boundary Lagrangian and $K_{ab}$ is the extrinsic curvature of the boundary.
Since the boundary (\ref{boundary}) is unidimensional the only possible boundary Lagrangians are
either of a point particle or a worldsheet of mass $\mu$ for a timelike or spacelike worldline.
As we show in the appendix, the variation of the boundary action (\ref{boundary}) is trivial. We therefore have the choice
of setting the boundary action to vanish on the semiclassical solution. In this way the boundary term will be
irrelevant in the semiclassical calculations and therefore the techniques used in \cite{san} straightforwardly
applies to our case.

In the semiclassical approximation this theory is described by the set of equations
\begin{subequations} \label{equ1}
\begin{align}
\label{e1} &\Lambda g_{ab}=\langle T_{ab}\rangle\ , \\ \label{bocon} &K=\mu\ ,
\end{align}
\end{subequations}
where we consider negligible the quantum correction to the boundary Lagrangian.

The trace anomaly of the conformal field theory, can be determined by the only knowledge of the background
geometry and it is \cite{BD}
\be \label{CA}
 \langle T^a{}_a\rangle=-\frac{\hbar\gamma}{24\pi}R\ .
\ee
$\gamma$ is proportional to the number of fields in the theory where matter fields are counted with opposite signs with
respect to the graviton contribution.

Using the gauge freedom in fixing the coordinates we can write the spacetime metric as
\begin{equation}\label{metrsun}
ds^2=-\Omega^2(u,v)dudv \ .
\end{equation}
Conservation equations $\nabla^b\langle T^a{}_b\rangle=0$, equations \eqref{e1} and \eqref{CA}
give the following equations
\begin{subequations}\label{eq}
\begin{align}
-\frac{1}{2}\Lambda \Omega^2&=\langle T_{uv}\rangle\ , \\
 0&=\langle
T_{uu}\rangle=\langle T_{vv}\rangle\ , \\
\langle T_{uu}\rangle&=-\frac{\gamma}{12\pi}\Omega\partial_u^2 \Omega^{-1}+ \tilde{U}(u)\ , \\ \langle
T_{vv}\rangle&=-\frac{\gamma}{12\pi}\Omega\partial_v^2 \Omega^{-1}+ \tilde{V}(v)\ , \\ \hbar^{-1}\langle
T^a{}_a\rangle&=-\frac{\gamma}{24\pi}R \ ,
\end{align}
\end{subequations}
where $\tilde U$ and $\tilde V$ set the vacuum state in which equation (\ref{se2}) is solved.

We now set $\tilde U$ and $\tilde V$ constant and proportional to the number of fields $\gamma$
by writing
\be \label{eq2}
\tilde{U}=\tilde{V}=\frac{q}{48\pi} \gamma\ ,
\ee
where $q$ is a constant. As we will see later this choice corresponds to setting the vacuum to be the
Hartle-Hawking state.

Equations (\ref{eq}) can now be solved for $\Omega$ and give
\be\label{cfnull}
\Omega^2=\frac{4 q}{\lambda^2} \frac{e^{(v-u) \sqrt{q}}}{(1+e^{(v-u)
    \sqrt{q}})^2}\ ,
\ee
where
\be \label{l21}
\lambda^2=\frac{48\pi\Lambda}{\hbar\gamma}\ .
\ee

In order to understand the physical meaning of the two constants $q$ and $\lambda$ we rewrite
our metric in the Schwarzschild gauge
\be\label{met}
ds^2=-f(x)dt^2+\frac{dx^2}{f(x)}
\ee
by setting
\begin{align*}
&q=\lambda^2 N+ M^2\ , \\
&t=\frac{v+u}{2}\
\end{align*}
and
\begin{subequations}
\begin{align}
&x_+=-\frac{\sqrt{q}}{\lambda^2}\frac{(1-e^{(v-u) \sqrt{q}})}{(1+e^{(v-u)
    \sqrt{q}})}+\frac{M}{\lambda^2}\ , \\
&f_+(x_+)=\lambda^2 x_+^2+2 M x_+-N
\end{align}
\end{subequations}
with
\be \label{xp}
M<x_+\lambda^2<\sqrt{q}+ M\ .
\ee
Alternatively, we can use, in place of $x_+$, the following coordinate
\begin{subequations}
\begin{align}
&x_-=-\frac{\sqrt{q}}{\lambda^2}\frac{(1-e^{(v-u) \sqrt{q}})}{(1+e^{(v-u)
    \sqrt{q}})}-\frac{ M}{\lambda^2}\ , \\
&f_-(x_-)=\lambda^2 x_-^2-2 M x_--N
\end{align}
\end{subequations}
with
\be \label{inter}
- M<x_-\lambda^2<\sqrt{q}- M \ .
\ee
The interval \eqref{inter} can also be restricted to
\be\label{xm}
- M<x\lambda^2<0\ .
\ee
We can analytically extend the ranges of values for $x_+$ in \eqref{xp} and $x_-$ in \eqref{xm} to
$0\le x_+<\infty$ and $-\infty<x_-\le 0$ respectively. We now implement the boundary condition \eqref{bocon} and we set
the boundary at $x=0$.
The manifold satisfying the boundary conditions \eqref{bocon} can be constructed by matching the two patches $x_+$ and
$x_-$ in $x_+=x_-=0$ and defining
\be\label{f}
f(x)=\lambda^2 x^2+2 M |x|-N\ ,
\ee
where $-\infty<x<\infty$ and $\mu=M/\sqrt{N}$ and so,
for a positive mass $\mu$,
it follows that $M$ and $N$ must be both non negative.
The black hole horizon is at
\begin{equation}\label{bhh}
    |x_{h}|=\frac{-M+\sqrt{q}}{\lambda^2} \ .
\end{equation}
and it is purely quantum. In fact, using equation \eqref{l21}, \eqref{bhh} becomes
\begin{equation}\label{xhh}
 |x_h|= \hbar \frac{\gamma}{48 \pi \Lambda} (-M+\sqrt{q}) \ ,
\end{equation}
and $\lim_{\hbar \rightarrow 0}x_h=0$.

Rescaling the coordinates, by defining $\tilde{t}=\sqrt{N} t$ and $\tilde{x}=x/\sqrt{N}$, we can rewrite our metric as
\be \nonumber
ds^2=-\tilde f(\tilde{x})d\tilde{t}^2+\frac{d\tilde{x}^2}{\tilde{f}(\tilde{x})}
\ee
with
\be \nonumber
\tilde{f}(\tilde{x})=\lambda^2 \tilde{x}^2+2 \mu |\tilde{x}|-1\ .
\ee
It is now clear that the metric depends only on the two physical quantities $\lambda^2$ and $\mu$ and
it is asymptotically $AdS$ with $AdS $ length $l=1/\lambda$. The black hole mass, the
spacetime mass when the $AdS$ contribution is subtracted, is therefore only determined
by the boundary mass and it is $E=\mu$ \footnote{In two dimensions there is no gravitational binding energy
contributing to the mass.}.

The black hole metric \eqref{met} with \eqref{f} was firstly found in \cite{mann} (where, due to different
boundary conditions, the
boundary mass is $\mu=M$ ) as solution of a different
two-dimensional gravity theory \cite{JT} and also here the black hole mass is
proportional to the mass
on the boundary \footnote{Other similar
solutions in two-dimensional dilatonic theory of gravity are considered in \cite{mariano2,mariano3}.}.

\subsection{Conformal properties}
In this section we explore the conformal properties of our solution.
Although the $(\tilde x,\tilde t)$ coordinates seem more natural, we will continue to use $(x,t)$
coordinates because, as we will see later, these are the one in which the correspondence with the
BTZ brane black hole is clearly manifest.

\begin{figure}[t]
\centering \includegraphics[angle=0,width=2in] {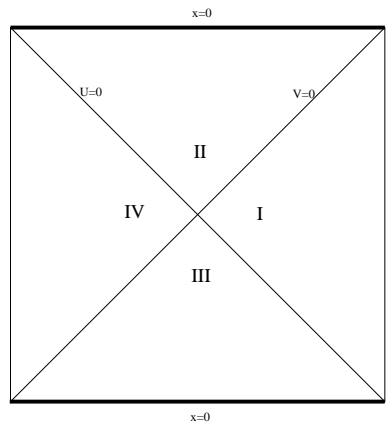}
\caption{ \label{uno} Conformal diagram of the
maximally extended black hole solution of metric \eqref{met} with \eqref{f}. The lines in bold represent
the boundary $x=0$.}
\end{figure}

Being $x=0$ a physical boundary we can just consider the patch $0< x < \infty$.
The black hole horizon is defined in \eqref{bhh}. We introduce a tortoise coordinate $r_*$ as
\begin{equation}
\begin{split}
  r_*&=\int{\frac{dx}{\lambda^2 x^2+2M x-N}} \\
  &=\frac{1}{2\sqrt{q}}\ln\left(\left|
  \frac{\sqrt{q}-\lambda^2 x - M}{\sqrt{q}+\lambda^2 x + M}\right
    |\right) \ ,
\end{split}
\end{equation}
and the horizon is now moved to  $r_* \rightarrow \infty$. Introducing null coordinates $u$ and $v$, such that
\begin{align}\label{nullc}
    &u=t-r_* \ , \\ \label{nullc2} &v=t+r_* \ ,
\end{align}
the conformal factor becomes the one in equation \eqref{cfnull}.
Note that the null coordinates $u$ and $v$ cover the full spacetime ($-\infty < u, v <\infty$) and at the
horizon $u\rightarrow \infty$ and $v\rightarrow -\infty$. These coordinates do not anyway represent a continuous
and complete set of coordinates across the horizon. In order to have a global set of coordinates we introduce
\begin{align}\label{Kru1}
    &U=-\frac{1}{\sqrt{q}}e^{-\sqrt{q}u}  \ , \\
    \label{Kru2} &V=\frac{1}{\sqrt{q}}e^{\sqrt{q}v} \ ,
\end{align}
and the metric becomes
\begin{equation}\label{cfkru}
    ds^2=-\frac{4 q}{\lambda^2}\frac{1}{1-qUV} dU dV \ .
\end{equation}
In this coordinate system the horizon is at $U\rightarrow 0$ and $V\rightarrow 0$ and  $-\infty<U<0$ and $0<V<\infty$.
The spacetime can be now analytically extended to the whole plane $-\infty < U, V <\infty$. We can also define the new
cartesian coordinates
\begin{align}\label{nullcb}
    &T=\frac{U+V}{2} \ , \\ &R=\frac{V-U}{2} \ .
\end{align}
The Penrose diagram of this maximally extended spacetime is shown in
fig. \ref{uno}.

\subsection{Temperature}

In this section we show that at our black hole is associated a physical quantum temperature due to the presence
of a boundary in $x=0$. Similar features are studied in \cite{mariano} for the dilatonic case.

We consider the quantization of a massless scalar field $\phi$ in
our two-dimensional spacetime, in this section we will use units such that $\hbar=1$. The wave equation
\begin{equation*}
    \square \phi =0
\end{equation*}
has solutions, with respect to the extended coordinates defined in
\eqref{Kru1} and \eqref{Kru2}, given by the orthonormal modes
\begin{equation*}
    \Phi_k=\frac{1}{\sqrt{4\pi\omega}}e^{i(k X-\omega T)} \ ,
\end{equation*}
where $\omega = |k|>0$ and $-\infty < k< \infty$. These modes are
positive frequency with respect to the timelike Killing vector
$\partial_T $, in fact they satisfy
\begin{equation*}
    \pounds_{\partial_T}\Phi_k=-i\omega\Phi_k \ .
\end{equation*}
The modes with $k>0$ consist of right-moving waves
$(4\pi\omega)^{-\frac{1}{2}}e^{-i\omega U}$ along the rays
$U=$constant, and they are analytic functions of $U$ and bounded in
the upper-half $V$-plane. The modes with $k<0$ consist of
left-moving waves $(4\pi\omega)^{-\frac{1}{2}}e^{-i\omega V}$along
the rays $V=$constant, and are analytic functions of $U$ and bounded
in the lower-half $U$-plane.

The general solution of the wave equation may be expanded as
\begin{equation}\label{expans1}
    \phi=\sum^{\infty}_{k=-\infty}(a_k \Phi_k+\a_{k}^{\dag}\Phi^*_k)
    \ .
\end{equation}
Upon quantization, $a_k$ and $\a_{k}^{\dag}$ become annihilation and creation operators
and the vacuum state for
the inertial observer is defined, as usual, by
\begin{equation}\label{vac1}
    a_k|0_A\rangle=0 \ .
\end{equation}
We can now also adopt an alternative quantization prescription based
on modes defined using the null coordinates of equations
\eqref{nullc},\eqref{nullc2}. The wave equation is conformally
invariant and we have mode solutions given by
\begin{equation} \label{modenull}
    \phi_k=\frac{1}{\sqrt{4\pi\omega}}e^{i(k x\pm\omega t)} \ ,
\end{equation}
where $\omega = |k|>0$ and $-\infty < k< \infty$. The upper sign in
\eqref{modenull} applies in region IV of the Penrose diagram in fig. \ref{uno}
and the lower sign in region I. The presence of this sign change
can be regarded as due to the fact that a right-moving wave in
region I moves towards increasing values of $x$, while in region IV
it moves towards decreasing  values of $x$, or simply due to the
time reversal we have in region IV. These modes are positive
frequency modes with respect the timelike Killing vector
$\partial_t$ in region I and $-\partial_t$ in region IV, satisfying
\begin{equation*}
    \pounds_{\pm\partial_t}\phi_k=-i\omega\phi_k \ ,
\end{equation*}
in region I and IV respectively.

We can now define
\begin{equation}\label{modes11}
    \phi^{(1)}_k=\left\{
                        \begin{array}{ll}
                          \frac{1}{\sqrt{4\pi\omega}}
                          e^{i(k x-\omega t)}, & \hbox{in region I;} \\
                          0, & \hbox{in region IV,}
                        \end{array}
                      \right.
\end{equation}
and
\begin{equation}\label{modes22}
     \phi^{(4)}_k=\left\{
                        \begin{array}{ll}
                          0, & \hbox{in region I;} \\
                          \frac{1}{\sqrt{4\pi\omega}}
                          e^{i(k x+\omega t)}, & \hbox{in region IV.}
                        \end{array}
                      \right.
\end{equation}
The set in equation \eqref{modes11} is complete in region I while the set in \eqref{modes22} is complete in
region IV, but neither set separately is complete on all our spacetime. However both sets together are
complete, and lines $t=$constant taken across both region I and IV are Cauchy surfaces for the whole spacetime.
Therefore, these modes can also be analytically continued into the regions II and III and so can be used as a
based for quantizing the field $\phi$ that can be then expanded as
\begin{equation}\label{expans2}
    \phi=\sum^{\infty}_{k=-\infty}(b^{(1)}_k \phi^{(1)}_k+
    b^{(1)\dag}_k \phi^{(1)*}_k+b^{(2)}_k \phi^{(4)}_k+b^{(2)\dag}_k \phi^{(4)*}_k)
    \ ,
\end{equation}
and the vacuum state can be now defined as the one satisfying
\begin{equation}\label{vac2}
    b^{(1)}_k|0_B\rangle=b^{(2)}_k|0_B\rangle=0 \ .
\end{equation}
This vacuum state is obviously not equivalent to the one defined in
\eqref{vac1} as we can easily see by analyzing the different modes.
To derive the Bogolubov transformations relating the operators
$b^{(1)}_k$ and $b^{(2)}_k$ to the operators $a_k$ of the inertial
observer we will follow an argument due to Unruh \cite{Unruh1}.

Note that the solution \eqref{modes11} with support in region I can
be extended to region II and the solution \eqref{modes22} with
support in region IV can be extended to region III and we can define
the new following modes
\begin{align}\label{psi1}
    &\psi^{(1)}_k=\phi^{(1)}_k+e^{-\pi \omega /
    \sqrt{q}}\phi^{(4)*}_{-k} \ ,\\
    \label{psi2}
    &\psi^{(4)}_k=\phi^{(1)*}_{-k}+e^{\pi \omega /
    \sqrt{q}}\phi^{(4)}_k \ ,
\end{align}
that are all defined in the entire spacetime (all four regions) and that represent a set of positive-energy
solutions of the wave equation. Therefore, an inertial observer may expand a general solution as
\begin{equation}\label{expans3}
    \phi=\sum^{\infty}_{k=-\infty}(C^{(1)}_k \psi^{(1)}_k+
    C^{(1)\dag}_k \psi^{(1)*}_k+C^{(2)}_k \psi^{(4)}_k+C^{(2)\dag}_k \psi^{(4)*}_k)
    \ ,
\end{equation}
and the vacuum state can be now defined as the one satisfying
\begin{equation}\label{vac3}
    C^{(1)}_k|0_A\rangle=C^{(2)}_k|0_A\rangle=0 \ .
\end{equation}

We can now easily relate these modes to the $b^{(1)}_k$ and
$b^{(2)}_k$ by using equations \eqref{psi1}, \eqref{psi2} and
\eqref{expans3} and we obtain
\begin{align}\label{Bog1}
 & b^{(1)}_k=C^{(1)}_k+e^{-\pi \omega /\sqrt{q}}C^{(2)\dag}_k\\
 & b^{(2)}_k=C^{(2)}_k+e^{-\pi \omega /\sqrt{q}}C^{(1)\dag}_k
\end{align}
The $C$-modes are not properly normalized. From the commutation
relations for the $b$-modes
\begin{equation}\nonumber
    \left[b^{(r)}_k,b^{(s)\dag}_{k'}\right]= \delta^{rs} \delta_{kk'}
\end{equation}
we deduce
\begin{equation}\nonumber
    \left[C^{(r)}_k,C^{(s)\dag}_{k'}\right]= \frac{e^{\pi \omega /
    \sqrt{q}}}{2\sinh(\pi \omega / \sqrt{q})}\delta^{rs} \delta_{kk'}
\end{equation}
and so we define the normalized creation and annihilation operators
by
\begin{equation}\nonumber
    c^{(r)}_k=e^{-\pi \omega /\sqrt{q}} \sqrt{2\sinh(\pi \omega / \sqrt{q})} C^{(r)}_k
\end{equation}
so that
\begin{equation}\nonumber
    \left[c^{(r)}_k,c^{(s)\dag}_{k'}\right]= \delta^{rs} \delta_{kk'}
\end{equation}
The $b^{(r)}_k$ operators can now be written in terms of the
$c^{(r)}_k$ as follows
\begin{align}\label{Bogv}
 &b^{(1)}_k= \frac{1}{\sqrt{2\sinh(\pi \omega / \sqrt{q})}} \left\{e^{\pi \omega /
 2\sqrt{q}} c^{(1)}_k + e^{-\pi \omega / 2\sqrt{q}} c^{(2)\dag}_k
 \right\} \ ,
 \\ \label{Bogvb}
 &b^{(2)}_k= \frac{1}{\sqrt{2\sinh(\pi \omega / \sqrt{q})}} \left\{e^{\pi \omega /
 2\sqrt{q}} c^{(2)}_k + e^{-\pi \omega / 2\sqrt{q}} c^{(1)\dag}_k
 \right\}
\end{align}
and these are the Bogolubov transformation relating the states
$|0_A\rangle$ and $|0_B\rangle$.

Now suppose the system is the state $|0_A\rangle$, the number
operator for the observer associated to $|0_B\rangle $, is simply
given by
\begin{equation}\nonumber
    N(k)=b^{(1)\dag}_k b^{(1)}_k
\end{equation}
since $b^{(2)\dag}_k$ excites modes which vanish in region I and are
therefore non accessible to the observer whose trajectory is in
region I. Using the Bogolubov transformation \eqref{Bogv},
\eqref{Bogvb} and the definition \eqref{vac3} of the vacuum state
$|0_A\rangle$, we obtain the expectation value of the number
operator
\begin{equation}\nonumber
   \langle 0_A|N(k)|0_A\rangle=\frac{e^{-\pi \omega /\sqrt{q}}}{2\sinh(\pi \omega /
   \sqrt{q})}=\frac{1}{e^{2\pi \omega /\sqrt{q}}-1} \ ,
\end{equation}
and this is precisely the Planck spectrum for radiation at
temperature, replacing $\hbar$,
\begin{equation}\label{tempe}
    T=\hbar \frac{\sqrt{q}}{2\pi k_B}
\end{equation}
where $k_B$ is the Boltzman constant.
Note that we get the same result for the
temperature by considering the Wick rotation in imaginary time as it was done in \cite{mann} for the black hole
of the two dimensional JT gravity theory \cite{JT}.

The heat capacity of our black hole is given by
\be
C=\frac{d \mu}{d T}=\frac{4\pi^2 k_B^2}{\mu\hbar N}T\ ,
\ee
We can see that the black hole has a positive heat capacity and therefore
it can reach thermal equilibrium with the thermal bath due to the Hawking radiation.

From equations \eqref{eq} with \eqref{eq2} we have that the vacuum expectation value of the
normal ordered stress tensor operator is simply given by
\begin{align*}
&\langle :T_{uu}: \rangle =\tilde{U}=\frac{\gamma}{48\pi} q\ \\
&\langle :T_{vv}: \rangle =\tilde{V}=\frac{\gamma}{48\pi} q\
\end{align*}
after transforming to extended coordinates $U$ and $V$ via the Schwarzian derivative we get \cite{FN}
\begin{equation}\nonumber
\langle :T_{UU}: \rangle =\langle :T_{VV}: \rangle=0
\end{equation}
and so, as we stated before, our semiclassical equation are actually solved
in the Hartle-Hawking vacuum state \cite{HH}.

\section{A brane in BTZ}

As we previously discussed, the two dimensional black hole described above is purely quantum, by means that the
presence of the horizon is due only to quantum mechanical effects. The holographic conjecture of \cite{em,tan}
implies that boundaries of some asymptotically AdS spaces should correspond to our semiclassical solution. The
only known (asymptotically $AdS$) black hole in three dimensions is the BTZ one \cite{BTZ}. We then expect our
solution to be a slice of a BTZ black hole.

A boundary solution with non zero vacuum energy (equivalent to a UV cutoff on the brane) is equivalent to a
braneworld solution \cite{RS}. A braneworld is a slice (brane) of a given bulk once a $Z_2$ symmetry with respect to the
brane is introduced. In our case the system is governed by the following action
\be\label{ac}
\begin{split}
{\cal A}_g=\frac{1}{2k_3^2}\int
d^3 x (R+\frac{2}{L^2})\sqrt{-g}-\frac{2\sigma}{k_3^2}\int_{\Sigma} d^2 x\sqrt{-h}\ ,
\end{split}
\ee
where $n^\alpha$ is the normal to the brane $\Sigma$ and $L$ is the $AdS_3$ length;
$\sigma$ and $h_{\alpha\beta}$ represent
the vacuum energy of and the induced metric on the brane and $k_3^2$ is the inverse of
the three dimensional Planck mass.

The vanishing of the variation of the action (\ref{ac}) implies the Einstein equations
\begin{equation*}
    R_{\mu \nu}-\frac{1}{2}R g_{\mu \nu}=\frac{1}{L^2}g_{\mu \nu} \ ,
\end{equation*}
with the boundary condition \cite{israel}
\be\label{jc}
K_{\alpha\beta}=\sigma h_{\alpha\beta}\ ,
\ee
where $K$ is the extrinsic curvature of $\Sigma $.

\subsection{1-brane}

We want to introduce a static 1-brane in the 3D BTZ black hole spacetime. In order to do that we consider the
surface
\begin{equation}
\Sigma : \theta - \Psi(r)=0
\end{equation}
whose normal is given by
\begin{equation} \label{normal}
n_{\alpha}=\pm A(0,-\Psi',1)\ ,
\end{equation}
where $'=\partial_r$ and the normalization factor $A=r(\Psi' r^2 F(r)+1)^{-1/2}$. The $\pm$ sign is related to
the orientation of $\Sigma$ as we shall see later.

Equation (\ref{jc}) is solved, in the BTZ background, by the function $\Psi(r)$
\begin{eqnarray}\label{psi}
\Psi_{\pm}(r)=\pm\frac{\ln \left( \frac{2 \sigma ^2 L^4 m +2\sigma
L^2\sqrt{m}\sqrt{r^2(1-\sigma^2L^2)+\sigma^2L^4 m} }{L r} \right)}{\sqrt{m}}
\end{eqnarray}
As we can see there are two different branches of solution. We will call these two branches the $+$ and $-$ branch and
we will indicate them  with $\Psi_+$ and $\Psi_-$ respectively.
The periodicity condition of $\theta$ in (\ref{F}) implies
\begin{equation} \nonumber
\Psi_\pm(r) \equiv \Psi_\pm(r) + 2\pi\ .
\end{equation}
To have a lighter notation we introduce the following quantities
\begin{equation}\label{alebe}
    \alpha=2\sigma^2 L^3 m \qquad \beta=4\sigma^2 L^2 m
    (1-\sigma^2 L^2)\ ,
\end{equation}
so that
\begin{equation} \nonumber
\Psi_\pm(r)=\pm \frac{1}{\sqrt{m}} \ln \left( \frac{\alpha +\sqrt{\beta r^2+ \alpha^2} }{r} \right)\ .
\end{equation}

The induced metric on the brane is given by
\begin{equation}\label{ind}
ds^2_\pm=-\left(\frac{r^2}{L^2}-m \right)dt^2+\left(\frac{1}{\frac{r^2}{L^2}-m}+r^2 \Psi'^2_\pm\right) dr^2\ ,
\end{equation}
and so the Ricci scalar is given by
\begin{equation}\nonumber
    R_\pm=-\frac{2m\beta}{m L^2\beta+\alpha^2}=-\frac{2}{L^2}(1-\sigma^2L^2)\ .
\end{equation}
We can easily see that the two dimensional brane is indeed asymptotically AdS (if $\sigma^2 \ne L^{-2}$)
with cosmological constant
\begin{equation}\nonumber
    \Lambda_2=-\frac{1}{L^2}(1-\sigma^2 L^2)\ .
\end{equation}

We now turn our attention to the properties of the slice and we consider only the $+$ branch
$\Psi_+(r)$ of (\ref{psi})
that from now on we will simply indicate as $\Psi(r)$ (the analysis of the $-$ branch is
completely analogous to the following
one). We have
\begin{equation}\nonumber
    \Psi'=\frac{\partial \Psi}{\partial
    r}=-\frac{\alpha}{\sqrt{\beta r^2+\alpha^2} \sqrt{m} r} <0\ ,
\end{equation}
so our function $\Psi(r)$ is always decreasing and also
\begin{equation}\label{rpsiimde}
  (r^2 \Psi'^2)'=-\frac{2\alpha^2\beta r}{(\beta r^2+\alpha^2)^2 m}\ .
\end{equation}
The right hand side of (\ref{rpsiimde}) is equal to zero in $r=0$ and always decreasing after that. Considering
this, since
\begin{equation}\label{grrim0}
    h_{tt}(0)h_{rr}(0)=-1<0\ ,
\end{equation}
we have that $h_{tt}h_{rr}<0$ always, avoiding Euclidean patches on the brane.

\begin{figure}[t]
\centering\includegraphics[angle=0,width=3.2in]{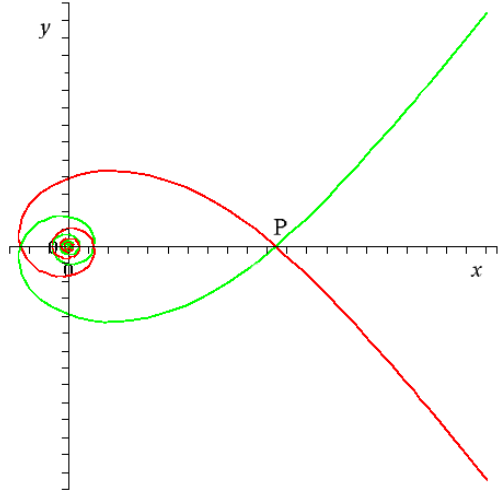}
\caption{Plot of $\Psi_+$ (red line) and $\Psi_-$ (green line) in a cartesian plane $(x,y)$,
$x=r\cos\theta$ and $y=r\sin\theta$ with $\theta=\Psi_{\pm}$. $P$ is the point $(r_{n_{\hbox{\tiny max}}},0)$ and it
is the last point in which the two branches intersect.
\label{wrap}}
\end{figure}

Given the periodicity of $\Psi$ we are now interested in the points $r=r_n$ in which the brane makes a full
loop (\textit{i.e.} where the two branches intersect on the cartesian $x$ axis, see fig. \ref{wrap}).
These points are defined by the equation $$\Psi(r=r_n)=n\pi$$ with $n$ integer. A solution is
 $r_0=0$ and the others are
\begin{equation}\label{solnp}
r_n=\frac{2\alpha e^{n \pi \sqrt{m}}}{e^{2 n \pi \sqrt{m}}-\beta}
\end{equation}
Note that
\begin{equation}\nonumber
    \frac{d r_n}{d n}<0
\end{equation}
and $r_n$ blows up if exists an integer $n=n_c$ such that
\begin{equation}\nonumber
e^{2n_c\pi \sqrt{m}}=4 L^2 \sigma^2 m (1-\sigma^2 L^2)\ .
\end{equation}
if such an integer does not exists, the value of $r$ at which the two brane intersect is
 $r=r_{n_{\hbox{\tiny max}}}$ where
\begin{equation}\label{nmax}
    n_{\hbox{\tiny max}}=\left[\frac{1}{2\pi\sqrt{m}}
    \ln\left(4L^2\sigma^2 m (1-\sigma^2L^2)\right)\right]_{\rightarrow}>0\ ,
\end{equation}
and with $[a]_{\rightarrow}$ we mean the next integer after $a$ if $a$ is not an integer. So if
$\frac{1}{2\pi\sqrt{m}}\ln\left(4L^2\sigma^2 m (1-\sigma^2L^2)\right)$
is an integer the brane will wrap around an infinite number of times. This will also happen in
the asymptotically flat case in which $\sigma^2=L^{-2}$.
In the more likely case in which this is not an integer, the brane will wrap around only for a finite number of
times and will then reach infinity with a defined asymptotic angle as it is shown if fig \ref{wrap}.

\subsection{Black Hole}

The induced metric (\ref{ind}) does not represent yet a black hole,
the presence of the horizon is indeed only due to an accelerated coordinate system.
In fact (\ref{ind}) can be easily transformed to the $AdS_2$ metric
\cite{mariano}. In order to find a black hole solution, we consider a
positive mass $\mu$ localized on the brane
which acts as a boundary of the brane.
The global three-dimensional $Z_2$-symmetric solution implementing this scenario, will therefore be constructed
by considering the portion of the spacetime whose boundary is given by the two profiles
$\Psi_{\pm}$ from the last intersection point ($r=r_{n_{\mbox{\tiny max}}}$) in the increasing $r$ direction (see
fig. \ref{wrap}).
In doing this we need to choose the sign of the normal in equation \eqref{normal}.
We must impose that, in cartesian coordinates, the $n^y$ component of the normal is negative
and, since we have,
\begin{equation}\nonumber
    n^{y}=-\sin \theta n^r-r\cos \theta n^{\theta} \ ,
\end{equation}
with $\theta=\Psi(r)$,
we can just impose that
\begin{equation}\nonumber
    \lim_{r\rightarrow \infty} n^{y}<0 \Rightarrow \frac{\alpha}{\sqrt{\beta m}L^2}\sin(\theta_{\infty})<0
\end{equation}
where
\begin{equation}\nonumber
\begin{split}
\theta_{\infty}&=\lim_{r\rightarrow\infty} \Psi(r) \\
&=\frac{1}{2\sqrt{m}}\ln\left(4L^2\sigma^2 m (1-\sigma^2L^2)\right) \ ,
\end{split}
\end{equation}
and so the sign is related to the sign of $\sin \theta_{\infty} $. Note that when $\theta_{\infty}=n\pi$, where
 $n$ is an integer, the sign of the normal cannot be defined and this is complete agreement with what
 we saw above (see discussion after equation \eqref{nmax}).

The conical singularity formed by the intersection of $\Psi_+$ and $\Psi_-$ describe a
specelike particle sitting on our
1-brane. From the action point of view this is equivalent to add to the action (\ref{ac}) the boundary lagrangian
\eqref{boundary}.

To make the above discussion more concrete we will use a different coordinate gauge (we will again consider
only the $+$ branch). By making the coordinate change
\be \nonumber
\rho=\frac{\sqrt{r^2(1-\sigma^2 L^2)+\sigma^2L^4m}}{1-\sigma^2 L^2}\ ,
\ee
the transformed metric will verify the property $g_{tt}=g_{\rho\rho}^{-1}$.
As we said we would like to truncate the range of $r$ to be $r_{n_{\mbox{\tiny max}}}\leq r<\infty$. This implies a
minimum value for the new coordinate $\rho$, given by
\be \nonumber
\rho_m=\sqrt{\frac{r_{n_{\mbox{\tiny max}}}^2}{1-\sigma^2
L^2}+\frac{\sigma^2 L^4 m}{(1-\sigma^2 L^2)^2}}\ .
\ee
We now shift this point to the origin by setting
\be \nonumber
x=\rho-\rho_m\ ,
\ee
so that $0\leq x<\infty$. We now copy and paste the branch $\Psi_-$ in $x=0 $. Equivalently we extend $x$ to the range
$-\infty<x<\infty$ and we require that $g_{\alpha\beta}(-x)=g_{\alpha\beta}(x)$. By setting
\be \label{l22}
\lambda^2=\frac{1-\sigma^2 L^2}{L^2}\ ,
\ee
and
\begin{eqnarray} \nonumber
M&=&\frac{1}{L}\sqrt{\frac{r_{n_{\mbox{\tiny max}}}^2}{L^2}(1-\sigma^2
L^2)+\sigma^2 L^2 m}\ ,\\ \nonumber N&=& m-\frac{r_{n_{\mbox{\tiny max}}}^2}{L^2}\ ,
\end{eqnarray}
we obtain the induced metric on the brane to be
\be
\begin{split}\label{slice}
ds^2=&-\left(\lambda^2 x^2+2M|x|-N\right)dt^2 \\
&+\frac{dx^2}{\left(\lambda^2 x^2+2M|x|-N\right)}\ .
\end{split}
\ee
The metric (\ref{slice}) is equivalent to the
metric (\ref{met}) with the function $f$ given by (\ref{f}) and therefore represent a black hole surrounding a
boundary mass $\mu$.

Given that boundary mass $\mu=M/\sqrt{N}$ is not negative, we find again that $N>0$. This condition now implies that
our brane must cross the BTZ horizon and therefore that brane and bulk black hole must share the same horizon. In fact
from the condition $N>0$ we get that
\begin{equation}\label{oriz}
r_{n_{\hbox{\tiny max}}}<r_h=\sqrt{m}L\ ,
\end{equation}
so
\begin{equation}\nonumber
\frac{2 \alpha e^{{n_{\hbox{\tiny max}}} \pi \sqrt{m}}}{e^{2 {n_{\hbox{\tiny max}}} \pi \sqrt{m}}-\beta}<\sqrt{m}L\ .
\end{equation}
By setting $x=e^{{n_{\hbox{\tiny max}}}\pi\sqrt{m}}$ we have that
\begin{equation} \nonumber
x-\left(\frac{x^2}{4L^2\sigma^2 m}-(1-\sigma^2 L^2)\right)\sqrt{m} <0\ ,
\end{equation}
and so being $x>0$ we need
\begin{equation} \nonumber
x>2\sqrt{m}L \sigma (1+L \sigma)\ ,
\end{equation}
or
\begin{equation}\label{n>}
{n_{\hbox{\tiny max}}}>\frac{\ln(2\sqrt{m}L \sigma (1+L \sigma))}{\pi \sqrt{m}}\ .
\end{equation}
We can always write
\begin{equation} \nonumber
{n_{\hbox{\tiny max}}}= 1+\frac{1}{2\pi\sqrt{m}}\ln\left(4L^2\sigma^2 m (1-\sigma^2L^2)\right) -\epsilon
\end{equation}
where $0<\epsilon<1$. With this (\ref{n>}) reduces to
\begin{equation} \nonumber
e^{2(1-\epsilon)\pi\sqrt{m}} \frac{1-\sigma L}{1+\sigma L}>1\ .
\end{equation}
As $\frac{1-\sigma L}{1+\sigma L}<1$, to satisfy \eqref{oriz} we need a massive enough BTZ black hole. This is in line with the discussion
of \cite{Witten} which require the bulk black hole to have a large mass in order to be quantum mechanically
stable and to correspond to a $CFT$ in thermal equilibrium.

\section{Conclusions}

Motivated by the conjectured duality between braneworld bulk black holes and semiclassical black holes
of \cite{em,tan} we studied two-dimensional quantum black holes and 1-brane slices of a three
dimensional BTZ bulk black hole.

We found a new static two-dimensional quantum black hole solution surrounding a boundary mass, in thermodynamical
equilibrium in the Hartle-Hawking vacuum state.
This solution exists only if the two-dimensional cosmological constant is non-zero, as the conformal field theory
relates the trace anomaly to the cosmological constant.

The proposed duality would imply the existence of a static asymptotically $AdS$ two-dimensional brane black hole with
non-zero tension as a slice of an asymptotically AdS thermodynamically stable \cite{Witten} three-dimensional space.
Studying slices of the BTZ black hole we found that indeed, for massive enough bulk black hole, such a solution does
exist only in the non-vanishing cosmological constant case and we showed that it shares the same
geometry of our quantum solution. We also found a resonance between
the BTZ parameters and the 1-brane tension for which such a construction is impossible. It would be interesting
to reinterpret it, in the holographic prospective, from the point of view of a deformed conformal
field theory living on the spacial infinity of BTZ, extending \cite{deformed} to the
finite temperature case, however this is beyond the scope of this paper.

In any case, we can go a bit further with the duality between the two black holes,
expressing the temperature of our two dimensional quantum black hole in
terms of the
parameter $M$ and $N$ obtained from the slicing of BTZ. We find that
\be\label{t}
T=\hbar\frac{\sqrt{q}}{2\pi k_B}=\hbar\frac{\sqrt{m}}{2\pi k_B L}\ .
\ee
The temperature (\ref{t}) is the same temperature of the bulk BTZ black hole \cite{vakkuri}.
The boundary theory must therefore be a $TCFT$ (with a UV cutoff) and temperature given by the bulk black hole as
we would expect \cite{Witten}. This also fix the choice of the time coordinate to be $t$ instead of $\tilde t$.
It therefore seems that the conjectured duality between
classical bulk black holes and quantum brane black holes \cite{em,tan} applies to our case.

In three dimensions the holographic relation \cite{mal} reads
\be \nonumber
\gamma=\frac{12\pi L}{k_3^2}>0\ .
\ee
As we said $\gamma$ is proportional to the sum of the number of matter fields and the number of gravitons.
Matter fields are counted positively
and gravitons negatively \cite{san}. It is then clear that the theory describing our black hole
has to be a matter dominated one.

Equating (\ref{l21}) with (\ref{l22}) we obtain
\be
\nonumber \Lambda=\hbar \frac{1-\sigma^2L^2}{\sigma L}>0\ .
\ee
The configuration $\Lambda>0$ and $\gamma>0$ cannot be obtained classically, indeed if $\Lambda>0$ (positive
energy density in the Universe) one expects, classically, to have a positive curvature (this is the case in
dilatonic gravity ). In our case instead, starting from a positive cosmological constant, we get a negative
curvature. This is one of the possibilities envisaged in \cite{san} absent in the classical theory.

The fact that our black hole solution is due to the presence of matter might imply that our solution should correspond
to the ending state of a gravitational collapse. A very interesting question is therefore if a classical collapse of
a brane can be also holographically described as a quantum gravitational collapse in the semiclassical theory
we considered, however this is beyond the scope of this paper and we leave it for future work.

\acknowledgments
\section*{Acknowledgments}
CG wish to thank Mariano Cadoni, Christophe Galfard and Robert Mann for useful comments.
CG wish also to thank Daniele Oriti
and Etera Livinie for discussion on the boundary actions.
GPP wish to thank Alessandro Fabbri and Stephen Hawking for useful discussions.
CG~is supported by PPARC research grant (PPA/P/S/2002/00208) and GPP~is supported
by PPARC and the Gates Cambridge Trust.

\appendix
\section{APPENDIX: Boundary action}

We here show that the variation of the boundary extrinsic curvature with respect to the
boundary induced metric is trivial. Consider in fact
\be \nonumber
\int_{\partial\Sigma}dt \sqrt{h}K\ ,
\ee
standard calculations show that its variation read (see for example \cite{davis})
\be \nonumber
\int_{\partial\Sigma}dt \sqrt{h}\left(K_{\alpha\beta}-h_{\alpha\beta}K\right)\delta h^{\alpha\beta}\ ,
\ee
and since in one dimension $K_{\alpha\beta}-h_{\alpha\beta}K\equiv 0$, the above variation is zero.
This result implies that in two dimensions the extrinsic curvature of a given boundary can be freely fixed to
a value $\mu$, where $\mu$ represent the mass associated with the boundary.
In particular, in this paper, we would like to interpret the boundary mass $\mu$ as the mass of a spacelike particle.

The action associated with a spacelike point-particle is
\be\label{act}
I_b=\mu\int d\tau \sqrt{u_\alpha u_\beta g^{\alpha\beta}}\ ,
\ee
where $dx^\alpha u_\alpha=d\tau$ is the proper length of the particle worldsheet and
$u^\alpha$ is the two-velocity of the particle. In two dimensions, the boundary metric
defined by the particle worldline is $h_{\alpha\beta}=u_\alpha u_\beta$.
Therefore the action (\ref{act}) can be rewritten as an
explicit boundary action
\be \nonumber
I_b=\mu \int_{\partial\Sigma} dt \sqrt{h}\sqrt{u_\alpha u_\beta h^{\alpha\beta}}\ ,
\ee
where $\mu>0$ is the positive boundary mass \cite{mann}
and $\partial\Sigma$ is the boundary defined by the particle worldsheet.

The variation of this action with respect to $\delta h^{\alpha\beta}$ is
\begin{equation} \nonumber
\begin{split}
\delta I_b=-\frac{\mu}{2}\int_{\partial\Sigma} dt \sqrt{h}
\delta h^{\alpha\beta} \left(  \frac{u_\alpha u_\beta}{\sqrt{u_\mu u_\nu h^{\mu\nu}}} \right.
\left. -h_{\alpha\beta}\sqrt{u_\mu u_\nu h^{\mu\nu}}\right)\ .
\end{split}
\end{equation}
Imposing now the normalization of the worldline vector $u^\alpha u_\alpha=1$ we find that the above variation is zero.
Therefore the only dynamical equation is obtained from the variation of $x^\alpha$, i.e.
the equation of motion of the particle worldsheet.
In order to interpret the boundary mass $\mu$ as the particle mass we therefore need to show that the particle can sit
on the boundary chosen, given a spacetime metric.

We consider our spacetime in the natural coordinates $(\tilde t,
\tilde x)$, so that \be\nonumber ds^2=-\tilde f(\tilde x)d\tilde
t^2+\frac{d\tilde x^2}{\tilde f(\tilde x)}\ , \ee where $\tilde
f(\tilde x)=\lambda^2 \tilde x^2+2\mu|\tilde x|-1$. The geodesic
equation is solved for \be\label{u} u^{\tilde t}=\frac{C}{\tilde f}\
,\ u^{\tilde x}=\sqrt{-\tilde f+C^2}\ , \ee where $C$ is the Energy
per unit mass of the particle. For physical reasons $|C|\geq 1$, as
the total energy of the particle cannot be smaller than the mass of
the particle itself. In particular the particle is at ``rest''
\footnote{In the spacelike region ($f<0$), at ``rest'' means
instantaneous.} for $|C|=1$. From (\ref{u}) we can therefore see
that the only point in which the particle is at ``rest'' is in $\tilde
x=0$, so that $u^{\tilde t}=1$ and $u^{\tilde x}=0$. This point is of an (unstable)
equilibrium as the potential $V=-\tilde f$ has a maximum in $\tilde x=0$.
We wish to comment here that since $\tilde x=0$ represent a point of
unstable equilibrium for the worldsheet,
the point particle approximation of a totally collapsed body can no
longer be used under perturbations and therefore, in this case,
a more detailed model for the collapsed matter has to be introduced
to study the stability of our system. However this study is beyond the
scope of the current paper and it is postponed for future research.

\end{document}